\documentclass[aps,floats,twocolumn,prd,nofootinbib,superscriptaddress]{revtex4-1}
\usepackage[utf8]{inputenc}
\usepackage{bm}
\usepackage{amsmath}
\usepackage{comment}
\usepackage{natbib, float, graphicx,amsmath,amsfonts,amssymb}
\usepackage[utf8]{inputenc}

\usepackage{amsmath}
\usepackage{graphicx}
\usepackage{dcolumn}
\usepackage{bm}
\usepackage[usenames, dvipsnames]{color}
\usepackage[normalem]{ulem}
\usepackage{xcolor}

\usepackage{scalerel,stackengine}
\stackMath
\newcommand\reallywidehat[1]{%
\savestack{\tmpbox}{\stretchto{%
  \scaleto{%
    \scalerel*[\widthof{\ensuremath{#1}}]{\kern-.6pt\bigwedge\kern-.6pt}%
    {\rule[-\textheight/2]{1ex}{\textheight}}%WIDTH-LIMITED BIG WEDGE
  }{\textheight}% 
}{0.5ex}}%
\stackon[1pt]{#1}{\tmpbox}%
}
\def\VEV#1{{\left\langle #1 \right\rangle}}

\newcommand{\SNR}{{\mathrm{SNR}}}

\def\cleb#1#2#3#4#5#6{\langle #3 \, #4 \, #5 \, #6 | #1#2 \rangle}
\def\ALMt#1#2#3#4{A^{#1 #2}_{#3 #4}}
\def\ALM{\ALMt{L}{M}{\ell}{\ell'}}

\begin{document}

\title{The search for statistical anisotropy in the gravitational-wave background with pulsar timing arrays}

\author{Selim~C.~Hotinli}
\affiliation{Astrophysics Group \& Imperial Centre for Inference
     and Cosmology, Department of Physics\\ 
     Imperial College London, Blackett Laboratory, Prince
     Consort Road, London SW7 2AZ, UK}

\author{Marc Kamionkowski}
\affiliation{Department of Physics of Physics and Astronomy, Johns Hopkins
     University, 3400 N.\ Charles St., Baltimore, MD 21218, USA}

\author{Andrew H.\ Jaffe}
\affiliation{Astrophysics Group \& Imperial Centre for Inference
     and Cosmology, Department of Physics\\ 
     Imperial College London, Blackett Laboratory, Prince
     Consort Road, London SW7 2AZ, UK}

\begin{abstract}
Pulsar-timing arrays (PTAs) are seeking gravitational waves from
supermassive-black-hole binaries, and there are prospects to
complement these searches with stellar-astrometry measurements.
Theorists still disagree,
however, as to whether the local gravitational-wave background
will be {statistically} isotropic, as arises if it is the summed
contributions from many SMBH binaries, or whether it exhibits
the type of {statistical} anisotropy that arises if the local
background is dominated by a handful (or even one) bright
source.  Here we derive, using bipolar spherical harmonics, the
optimal PTA estimators for {statistical} anisotropy in the 
GW background and simple estimates of the detectability of this
anisotropy.   We provide results on the smallest detectable amplitude of a dipole
anisotropy (and several other low-order multipole moments) and
also the smallest detectable amplitude of a ``beam'' of
gravitational waves.  Results
are presented as a function of the signal-to-noise with which
the GW signal is detected and as a function of the number of
pulsars (assuming uniform distribution on the sky and equal
sensitivity per pulsar).  We provide results first for
measurements with a single time-domain window function and then
show how the results are augmented with the inclusion of
time-domain information.  The approach here is intended to be
conceptually straightforward and to complement the results of
more detailed (but correspondingly less intuitive) modeling of
the actual measurements.
\end{abstract}

\maketitle

\section{Introduction}

A longstanding effort \cite{Foster:1990,Maggiore:1999vm,Burke-Spolaor:2015xpf,Lommen:2015gbz,
Hobbs:2017oam,Yunes:2013dva} to detect a stochastic
gravitational-wave background with pulsar-timing arrays consists
now of three major efforts---the Parkes Pulsar Timing Array (PPTA)
\cite{Hobbs:2013aka,Manchester:2012za}, North American Nanohertz
Observatory for Gravitational Waves (NANOGrav)
\cite{Arzoumanian:2018saf}, and the European Pulsar Timing Array
(EPTA) \cite{Lentati:2015qwp}--that collaborate through an
International Pulsar Timing Array (IPTA)
\cite{Verbiest:2016vem}.  The effects of gravitational waves on
the arrival times of pulses from pulsars
\cite{Detweiler:1979wn,Sazhin:1978} produce a
characteristic angular correlation \cite{Hellings:1983fr} in the
pulsar-timing residuals.  Signals at the frequencies
$\sim1$nHz are expected from the mergers
of supermassive-black-hole binaries
\cite{Rajagopal:1994zj,Jaffe:2002rt}.  There are also prospects
to use complementary information from stellar astrometry
\cite{Book:2010pf, Moore:2017ity, Mihaylov:2018uqm,
OBeirne:2018slh, Qin:2018yhy} as the apparent position of distant stars will oscillate 
with a characteristic pattern on the sky due to GWs.

It is still not understood, though, whether the local GW
signal due to SMBH mergers will be the type of stochastic
background that arises as the sum of a large number of
cosmological sources, or whether it will be dominated by just a
handful---or even just one---source \cite{Allen:1996gp,
Sesana:2008xk,Ravi:2012bz,Cornish:2013aba,Kelley:2017vox}.
Roughly speaking, if there are
$\sim N$ {Poisson} sources contributing to the signal, then the amplitude
of anisotropy in the GW background should be $\sim N^{-1/2}$.  A
first obvious step, after the initial detection of a
gravitational-wave signal, will therefore be to seek the
anisotropy in the background that may arise from a finite number
of sources.  Exotic sources might also lead to anisotropy
\cite{Kuroyanagi:2016ugi}.

Prior work \cite{Anholm:2008wy,Mingarelli:2013dsa,Gair:2014rwa} has developed
tools to characterize and seek with PTAs anisotropy in the GW
background that were then implemented in a null search
\cite{Taylor:2015udp}.  This anisotropy was characterized (as it
is here also) in terms of an uncorrelated and unpolarized
background of gravitational waves with a direction-dependent
intensity parametrized in terms of spherical-harmonic expansion
of the intensity. Here we re-derive anisotropy-detection tools using
mathematical objects (bipolar spherical harmonics; BiPoSHs
\cite{Hajian:2003qq,Hajian:2005jh,Joshi:2009mj})
developed for analogous problems in the study of the cosmic
microwave background.
The analysis here provides some simplifications and insights
and also intuitive estimates for the smallest detectable signals.
We provide numerical results for the smallest detectable
dipole-anisotropy amplitude as a function of the signal-to-noise
with which the isotropic signal is detected and as a function of
the number of pulsars in the array.  We restrict our attention
to PTAs but describe how the detectability will be augmented
with the inclusion of astrometry.

This paper is organized as follows:  In Section
\ref{sec:observables} we describe the idealized observables that
we model.  In Section \ref{sec:powerspectrum} we review the
standard Hellings-Downs correlation function (and its
harmonic-space equivalent, the timing-residual power spectrum)
used to detect the GW background.  Section \ref{sec:biposh}
introduces the bipolar-spherical-harmonic formalism and
describes how to infer the BiPoSH amplitudes from the
observables.  Section \ref{sec:model} describes the model
of an uncorrelated anisotropic background we consider here (and
considered in Refs.~\cite{Mingarelli:2013dsa,Gair:2014rwa}) and
calculates the BiPoSH coefficients for the model in terms of the
model's anisotropy parameters $g_{LM}$.  Section
\ref{sec:estimators} derives minimum-variance estimators for the
spherical-harmonic coefficients $g_{LM}$ that parametrize the
anisotropy and the variances $(\Delta g_{LM})^2$ with which they
can be measured.  Section \ref{sec:smallest} evaluates the
smallest detectable anisotropy beginning with a dipole and
then generalizing to anisotropies of higher-order multipole
moment and then the anisotropy due to a beam of uncorrelated
unpolarized gravitational waves from a specific direction.
Section \ref{sec:multiplemaps} describes how the previous
results, obtained for a single timing-residual map, are
generalized to incorporate the multiple maps that may be
obtained from time-domain information.  We discuss the extension
to astrometry and make closing remarks in Section
\ref{sec:discussion}.

\section{Harmonic and real-space angular observables}
\label{sec:observables}

PTA measurements are characterized by the temporal
evolution of the timing residuals and the dependence of
the observables as a function of position on the sky.  Here we
focus primarily on the angular structure.  To simplify, we
speak here of the ``timing residuals'' $z(\hat n)$ measured in a PTA as
a function of position $\hat n$ on the sky.  These
``timing residuals,'' in a more complete analysis, will be obtained
from some convolution of the timing residuals (TRs) with a
time-sequence window function (and there may well be a number of
such timing residuals that are obtained from convolutions of
the full timing-residual data with different time-sequence window
functions---more on this in Section~\ref{sec:multiplemaps}).
Strictly speaking, therefore, each appearance of
a GW power spectrum $P_h(k)$ in the expressions below should be
replaced by $P_h(k) \left[W(k)\right]^2$ where $W(k)$ is an
appropriate time-domain window function.

Any such timing residual $z(\hat n)$ can be expanded
\begin{equation}
     z(\hat n) = \sum_{\ell=2}^\infty \sum_{m=-\ell}^{\ell}
     z_{\ell m} Y_{\ell m}(\hat n),
\label{eqn:expansion}     
\end{equation}
in terms of spherical harmonics $Y_{\ell m}(\hat n)$, which
constitute a complete orthonormal basis for scalar functions on
the two-sphere.  The expansion coefficients are obtained from
the inverse transform,
\begin{equation}
     z_{\ell m} = \int\, d^2 \hat n\, z(\hat n) \, Y_{\ell
     m}^*(\hat n).
\label{eqn:inverse}     
\end{equation}
The sum in Eq.~(\ref{eqn:expansion}) is only over
$\ell\geq 2$, as the transverse-traceless gravitational waves
that propagate in general relativity give rise only to timing-residual
patterns with $\ell\geq 2$.  
We assume that the timing residuals (convolved with a time-sequence
window function) are real, and so $z_{\ell m}^* =(-1)^m z_{\ell,-m}$.
\footnote{In time-frequency Fourier space, $z(f)$ would be complex, but satisfy a similar reality condition.}
Note that specification of $z(\hat n)$ is equivalent to
specification of $z_{\ell m}$ and {\it vice versa}---they are
two different ways to describe the same observables.

\section{Power spectrum and correlation function}
\label{sec:powerspectrum}

The timing residuals $z({\hat n}, {\hat k})$ arising from a gravitational wave with polarization tensor $h_{ab}$ moving in direction ${\hat k}$ are given by 
\begin{equation}\label{eq:z_from_h}
     z(\hat n;\hat k) = \frac{n^a n^b h_{ab}}{2(1+\hat k \cdot
     \hat n)}\;.
\end{equation}
Strictly speaking, the timing residuals are observed as a function of time, but the angular pattern here is that after those time-domain data have been convolved with a time-domain window function so that the resulting map $z(\hat n)$ is then real. 

As discussed in Refs.~\cite{Roebber:2016jzl,Qin:2018yhy} (and below), the rotationally-invariant observed power spectrum $C_\ell \propto \sum_m |z_{\ell m}|^2/(2\ell+1)$ for this plane wave is 
\begin{equation}
     C_\ell \propto \frac{(\ell-2)!}{(\ell+2)!}.
\label{eqn:ell}
\end{equation}     
Since Eq.~(\ref{eq:z_from_h}) is a scalar and linear in $h_{ab}$, the timing residuals from {\it any} collection of plane waves---i.e., any gravitational-wave signal---will have the power spectrum of Eq.~(\ref{eqn:ell}).\footnote{It is mathematically possible---e.g., from a standing wave composed of two identical gravitational waves moving in opposite directions--- to get a different $\ell$ dependence, but hard to imagine how any astrophysical scenario could produce a power spectrum, that differs.} If the timing residuals $z(\hat n)$ arise from a realization of a statistically isotropic gravitational-wave background, then the spherical-harmonic coefficients $z_{\ell m}$ of the observed $z(\hat n)$ map will satisfy 
\begin{equation}
     \VEV{ z_{\ell m} z^*_{\ell'm'} } = C_\ell \, \delta_{\ell
     \ell'} \delta_{m m'},
\label{eqn:isopowerspectrum}
\end{equation}
where the angle brackets denote the average over all realizations of the gravitational-wave background, and $\delta_{\ell\ell'}$ and $\delta_{mm'}$ are Kronecker deltas. Eq.~(\ref{eqn:isopowerspectrum}) states that if the GW background is statistically isotropic then all of the $z_{\ell m}$ are uncorrelated and that each $z_{\ell m}$ is some number selected from a distribution of variance $C_{\ell}$. 
The resulting map, $z(\hat n)$, is then real after convolution
with the appropriate time-domain window function.

{The timing-residual power spectrum is related to the rotationally-invariant two-point}
autocorrelation function~\citep{Gair:2014rwa},
\begin{equation}
     C(\Theta) = \VEV{ z(\hat n) z(\hat m)}_{\hat n \cdot \hat m
     =  \cos\Theta} = \sum_\ell \frac{2 \ell+1}{4\pi} C_\ell
     P_\ell(\cos\Theta);
\label{eqn:autocorrelation}
\end{equation}
i.e., the product of the timing residuals in two different directions
separated by an angle $\Theta$, averaged over all such pairs of
directions.  The two-point autocorrelation function from a
stochastic GW background is the classic Hellings-Downs curve,
\begin{equation}
      C(\Theta) \propto (1/2)(1-x) \log \left[ \frac12 (1-x)
      \right] - \frac16 \left[ \frac12 (1-x)
      \right] +\frac13,
\end{equation}
where $x=\cos\Theta$. Again, the two-point autocorrelation function has this form
regardless of whether the GW background is statistically
isotropic or otherwise.

Since the power spectrum $C_\ell$ and two-point autocorrelation
function $C(\Theta)$ do not depend on whether the background is
isotropic or otherwise, the natural first step in any effort to
detect a GW background is to establish from the data that these
are nonzero.  Formulas to derive $C_\ell$ from (idealized) data
are provided below.

\section{Bipolar spherical harmonics}
\label{sec:biposh}

There is, however, far more information in a map $z(\hat n)$ (or
equivalently, its set of $z_{\ell m}$) than that provided by the
timing-residual power spectrum and Hellings-Downs correlation.  The most
general correlation between any two $z_{\ell m}$s can be
written (see, e.g., Ref.~\cite{Book:2011na,Pullen:2007tu}),
\begin{eqnarray}
     \VEV{ z_{\ell m} z^*_{\ell'm'}} &=& C_\ell \delta_{\ell\ell'}
     \delta_{mm'} \nonumber \\
     & & + \sum_{L=1}^\infty\sum_{M=-L}^L  (-1)^{m'}
     \cleb{L}{M}{\ell}{m}{\ell',}{-m'} \ALM,\nonumber \\
\label{eqn:bipoSHexp}
\end{eqnarray}
where $C_\ell$ is the (isotropic) power spectrum introduced
above, $\cleb{L}{M}{\ell}{m}{\ell'}{m'}$ are Clebsch-Gordan
coefficients, and the $\ALM$ are BiPoSH coefficients.  Note that
the power spectrum $C_\ell$ can be identified as
$(-1)^\ell A^{00}_{\ell\ell}/\sqrt{2\ell+1}$.

As Eq.~(\ref{eqn:autocorrelation}) indicates, the 
Hellings-Downs curve $C(\Theta)$ considers information obtained
only from the angular separation $\Theta$ between two directions
$\hat n$ and $\hat m$, but disregards any information about the
specific directions $\hat n$ and $\hat m$.  This additional
information is parametrized with BiPoSHs in terms of BiPoSH
coefficients $A^{LM}_{\ell \ell'}$ that characterize departures
from statistical isotropy.  If there is a dipolar power
anisotropy (higher flux of GWs from one direction than from the
opposite direction), it is characterized by the $L=1$ (dipolar)
BiPoSHs, and the different $M=0,\pm1$ components provide the
spherical-tensor representation of the dipole.  A quadrupolar
power asymmetry (e.g., as might arise if there were GWs coming
from the $\pm \hat z$ direction) are characterized by the $L=2$
BiPoSH coefficients, and so forth.

\subsection{Measurement of BiPoSH coefficients}

We suppose that the ``data'' come in the
form of a collection of measured values $z_{\ell m}^{\rm
data}=z_{\ell m} + z_{\ell m}^{\rm noise}$ each of which has a
contribution $z_{\ell m}$ from the signal and another $z_{\ell
m}^{\rm noise}$ from measurement noise.  We assume that the
noise in each $z_{\ell m}^{\rm noise}$ are uncorrelated and that
each $z_{\ell m}$ has a variance $N^{zz}$ (which we further assume to be 
$\ell$-independent --\,the white-noise power spectrum\,--
a good approximation if the timing-residual noises in all pulsars are comparable).

Estimators for the BiPoSH coefficients are then obtained from
\begin{equation}
     \reallywidehat{\ALM} = \sum_{mm'} z^{\rm data}_{lm} z^{*\,{\rm
     data}}_{l'm'} (-1)^{m'} \cleb{L}{M}{l}{m}{l',}{-m'}.
\label{eqn:ALMestimator}
\end{equation}
This estimator has a variance, under the null hypothesis (for
even $L+\ell+\ell'$) \cite{Book:2011na},
\begin{equation}
     \VEV{ \left|\reallywidehat{\ALM} \right|^2} = (1+\delta_{\ell\ell'})
     C_\ell^{\mathrm{data}} C_{\ell'}^{\mathrm{data}},
\label{eqn:ALMvariance}
\end{equation}
where $C_\ell^{\mathrm{data}} = C_\ell + N^{zz}$ is the power
spectrum of the map, which includes the signal and the noise.  The $\delta_{ll'}$ 
arises since the root-variance to a variance of a Gaussian distribution is $\sqrt{2}$ 
times the variance. We will see below that we need consider only combinations with
even $\ell+\ell'+L$.
If the map $z(\hat n)$ is real, then ${A^{LM}_{\ell\ell'} =
A^{LM}_{\ell'\ell}}$ (for even $\ell+\ell'+L$), and the
estimators $\reallywidehat{A^{LM}_{\ell\ell'}}$ and
$\reallywidehat{A^{LM}_{\ell'\ell}}$ are the same. The covariance
between any two other different $\reallywidehat{A^{LM}_{\ell\ell'}}$
vanishes.

By setting $L=0$ and identifying $C_\ell =
(-1)^\ell A^{00}_{\ell\ell}/\sqrt{2\ell+1}$, we recover the power-spectrum
estimator,
\begin{equation}
     \reallywidehat{C_\ell} = \sum_{m=-\ell}^\ell \frac{ |z_{\ell
     m}^{\rm data}|^2}{2\ell+1} - N^{zz},
\end{equation}
which has a variance
\begin{equation}
     \VEV{ \left(\Delta C_\ell \right)^2} = \frac{2}{2\ell+1} \left(C_\ell^{\rm
     data} \right)^2.
\end{equation}
{Under the null hypothesis of no gravitational-wave 
background (to be distinguished from the null hypothesis of a 
gravitational-wave background that is isotropic), $C_\ell^{\rm data}=N^{zz}$.  
This result will be used in Eq.~\eqref{eqn:SNR} below.}

\section{Model and BiPoSH Coefficients}
\label{sec:model}

We now focus on understanding the $\ell,\ell'$ dependence of the
BiPoSH coefficients $A^{LM}_{\ell\ell'}$.  To do so, we
must understand the dependence of the observable $z(\hat n)$ on
the gravitational-wave background.

\subsection{Model of anisotropic background}

{In order to link measurements of the timing residuals to an underlying 
gravitational wave background, we need a model for the statistics of that background.}
 Although there are an infinitude of ways the background can
depart from statistical isotropy, we consider (as did
Refs.~\cite{Mingarelli:2013dsa,Gair:2014rwa}) here those that
can be parametrized as
\begin{eqnarray}
     \VEV{ h_s(\vec k) h_{s'}^*(\vec k') } &=& \frac 14 \delta_{ss'}
     (2\pi)^3 \delta_D(\vec k -\vec k') P_h(k) \nonumber \\
     & & \times \left[1  +
     \sum_{L>0} \sum_{M=-L}^L g_{LM} Y_{LM}(\hat k)
     \right], \nonumber \\
\label{eqn:generalized}     
\end{eqnarray}
where $h_s(\vec k)$ is the amplitude of the gravitational-wave
mode with wavevector $\vec k$ and polarization $s=+,\times$.
With the Dirac delta function in this parametrization, we are
still preserving the assumption that different Fourier modes are
uncorrelated.  We are also assuming that the frequency
dependence of the GW background is the same in all
directions\footnote{This restriction is irrelevant,
given that the angular pattern induced by a gravitational wave
is independent of the GW frequency.}\ 
and that the $+$ and $\times$ modes are still equally
populated (i.e., that the background is unpolarized).  The sum
over spherical harmonics allows, however, for the most general
angular dependence of the gravitational-wave flux, parametrized
by spherical-harmonic coefficients $g_{LM}$.  Here, the
gravitational-wave power spectrum is $P_h(k)$, and an isotropic
background is recovered for $g_{LM}\to0$ for all {$L>0$}. {In this model, the $g_{LM}$ 
are the spherical harmonic coefficients of the map of total gravitational-wave power.}

Since the term in the brackets in Eq.~(\ref{eqn:generalized})
must be positive, the spherical-harmonic coefficients are
restricted to be $g_{L0} \leq \sqrt{4\pi/(2L+1)}$, and a roughly
similar bound applies to $\sqrt{2}\,\mathrm{Re} g_{LM}$ and
$\sqrt{2}\,\mathrm{Im} g_{LM}$ for $M\neq0$.

\subsection{Resulting timing-residual BiPoSH coefficients (and
angular power spectrum)}

We now calculate the BiPoSH amplitude that arises from a GW
background of the form in Eq.~(\ref{eqn:generalized}), based on its imprint, Eq.~(\ref{eq:z_from_h}).
If the GW direction is taken to be $\hat k =\hat z$, then this becomes
\begin{equation}
     z(\hat n; \hat k =\hat z) =h_+ (1-\cos\theta) \cos 2\phi +
     h_\times (1-\cos\theta) \sin 2\phi,
\end{equation}
where $h_+$ and $h_\times$ (both most generally complex) are the
amplitudes of the $+$ and $\times$ polarizations.

This plane wave is described by spherical-harmonic coefficients,
\begin{eqnarray}
     z_{\ell m}(\hat z) &=&z_\ell \left[ h_+ (\delta_{m2}+ \delta_{m,-2})
     + i h_\times (\delta_{m2}-\delta_{m,-2}) \right] \nonumber
     \\
     &=& z_\ell \left[ (h_+ + i h_\times) \delta_{m2} +
     (h_+-ih_\times) \delta_{m,-2}  \right], \nonumber \\
\label{eqn:zellm}     
\end{eqnarray}
where we defined
\begin{equation}
z_\ell \equiv (-1)^{\ell} \sqrt{\frac{{4\pi}(2\ell+1) (\ell-2)!}{(\ell+2)!}}\,.
\end{equation}  
From this result, we can construct the
spherical-harmonic coefficients for a plane wave in any other
direction.  To do so, we write
\begin{eqnarray}
     z(\hat n;\hat k)&=& \sum_{\ell m m'}  Y_{\ell m}(\hat n) D^{(l)}_{mm'}(\phi_k,\theta_k,0)
     z_{\ell m'}(\hat z)\,,
\end{eqnarray}
where $D^{(\ell)}_{mm'}(\phi_k,\theta_k,0)$ are
the Wigner rotation functions.\footnote{Strictly speaking,
this rotation most generally involves three Euler rotations.  We
will always choose, however, the $+$ and $\times$ polarizations
to align with the $\hat \theta$-$\hat \phi$ directions.  The
rotation thus involves first a rotation about the $\hat z$
direction by the azimuthal angle $\phi_k$ of $\hat
k=(\theta_k,\phi_k)$ and then another rotation by the polar
angle $\theta_k$.}  We thus infer, given
Eq.~(\ref{eqn:zellm}), which restricts the $m'$ sum to $\pm2$,
that a gravitational wave moving in the $\hat k$
direction imprints a pulsar-timing-residual pattern described by
spherical-harmonic coefficients,
\begin{eqnarray}
     z_{\ell m}(\hat k) &=& \sum_{m'}
     D^{(\ell)}_{mm'}(\phi_k,\theta_k,0) z_{\ell m'}(\hat z)
     \nonumber \\
     & = & z_\ell \left[ (h_+ + i h_\times) D^{(\ell)}_{m2} +
     (h_+ - i h_\times) D^{(\ell)}_{m,-2} \right],\nonumber \\
\end{eqnarray}
where $h_+$ and $h_\times$ are the GW amplitudes for this wave,
and the arguments of the rotation matrices are $(\phi_k,\theta_k,0)$.

Given this result, we can now calculate the BiPoSH coefficients
for a direction-dependent power spectrum of the form given in
Eq.~(\ref{eqn:generalized}).  We start by noting that a given
gravitational-wave pattern is described by a set of amplitudes
$h_+(\vec k)$ and $h_\times(\vec k)$ for each possible
wavevector $\vec k$.  The spherical-harmonic coefficients
induced by this gravitational-wave pattern are
\begin{equation}
     z_{\ell m}  = \sqrt{2} z_\ell \int \frac{d^3k}{(2\pi)^3} \left[
     h_R(\vec k) D^{(\ell)}_{m2} +h_L(\vec k) D^{(\ell)}_{m,-2}
     \right],
\end{equation}
where $h_R(\vec k) = 2^{-1/2}(h_++ i h_\times)(\vec k)$ and $h_L(\vec k)
= 2^{-1/2} (h_+ - i h_\times)(\vec k)$.  The correlation between any two
spherical-harmonic coefficients is therefore
\begin{eqnarray}
     \VEV{z_{\ell m} z_{\ell'm'}^*} &=& z_\ell z_{\ell'} \int
     \frac{d^3k}{(2\pi)^3} P_h(k) \left[ 1 + \sum_{LM} g_{LM}
     Y_{LM}(\hat k) \right]\nonumber \\
     & & \times \left[ D^{(\ell)}_{m2}(\hat k)
     \left(D^{(\ell')}_{m'2}(\hat k) \right)^* \right. \nonumber
     \\ & &  \left. \qquad + D^{(\ell)}_{m,-2}(\hat k)
     \left(D^{(\ell')}_{m',-2}(\hat k) \right)^* \right].
\label{eqn:withintegral}     
\end{eqnarray}
After performing the integral over directions $\hat k$ we find
an expression for $\VEV{z_{\ell m} z_{\ell'm'}^*}$ of the form
Eq.~(\ref{eqn:bipoSHexp}) with
\begin{equation}
     C_\ell = \frac{z_\ell^2}{2\ell+1} I,
\label{eqn:powerspectrum}     
\end{equation}
and
\begin{equation}
     A^{LM}_{\ell\ell'} = (-1)^{\ell-\ell'}(4\pi)^{-1/2} g_{LM} z_\ell
     z_{\ell'} H^L_{\ell\ell'} I,
\label{eqn:biposhresult}     
\end{equation}
where
\begin{equation}
     H^L_{\ell\ell'} \equiv \left( \begin{array}{ccc} \ell & \ell' &
     L \\ 2 & -2&  0 \end{array} \right),
\end{equation}
in terms of Wigner-3j symbols, and we defined ${I \equiv  [4\pi/(2\pi)^3] \int k^2\,dk\,
P_h(k)}$.  The two terms in Eq.~(\ref{eqn:withintegral}) cancel
if $\ell+\ell'+L$ is odd, and so $A^{LM}_{\ell\ell'}$ is nonzero
only for even $\ell+\ell'+L$.
There are two interesting features of
Eq.~(\ref{eqn:biposhresult}).  First, the
$\ell$ dependence appears only in the factors $z_\ell
z_{\ell'} H^L_{\ell\ell'}$;
as we will see below, this will allow us to write an
optimal estimator for the anisotropy coefficients $g_{LM}$.
Second, the power spectrum and BiPoSH coefficients both depend
in the same way on the power spectrum $P_h(k)$.

\section{Minimum-variance estimators of anisotropy}
\label{sec:estimators}

\subsection{Isotropic signal-to-noise}

Before evaluating the smallest detectable anisotropy, we write
the power spectrum in terms of the signal-to-noise ratio (SNR)
with which the isotropic signal is detected; this will be useful
below.  To do so, we recall that the variance with which any
given $C_\ell$ can be measured is $[2/(2\ell+1)] (N^{zz})^2$.
The signal-to-noise ratio SNR from a measurement that accesses
multipole moments up to $\ell_{\rm max}$ is then given by 
(see Ref.~\cite{Roebber:2016jzl} for a derivation)
\begin{equation}
 (\SNR)^2 = \sum_{\ell=2}^{\ell_{\rm max}} \frac{
 (2\ell+1)}{2} \left(\frac{C_\ell}{N^{zz}} \right)^2
\label{eqn:SNR}
\end{equation}    
and by using Eq.~(\ref{eqn:powerspectrum}), we find 
$I^2\simeq\,[1.17(\mathrm{SNR})N^{zz}]^2$ in the limit
$\ell_{\rm max}\to \infty$, which
turns out to be remarkably accurate for any finite $\ell_{\rm
max} \geq3$ given the very rapid decay of the summand with
$\ell$.  We can thus fix the gravitational-wave amplitude $I$ in
terms of the signal-to-noise ratio with which the isotropic
signal has been established, and the noise term, which will cancel in the estimator 
variance calculation below.\footnote{Eq.~(\protect\ref{eqn:SNR}) also indicates
that $\gtrsim93\%$ of the total signal to noise in the detection
of the GW signal comes from the quadrupole.}

\subsection{BiPoSH estimators and variance}

The observables that we seek to obtain from the data are the
anisotropy amplitudes $g_{LM}$.  Each estimator $\reallywidehat{
A^{LM}_{\ell \ell'}}$ provides an estimator,
\begin{equation}
    (\reallywidehat{g_{LM}})_{\ell \ell'} = (-1)^{\ell-\ell'}\sqrt{4\pi}\frac{
     \reallywidehat{A^{LM}_{\ell\ell'}}}{z_\ell z_{\ell'}
    H^L_{\ell \ell'} I},
\end{equation}
for $g_{LM}$.  The variance of each of these estimators is (for
$L+\ell+\ell'$ even),
\begin{eqnarray}
     (\Delta g_{LM})_{\ell\ell'}^2
     &=& \frac{4 \pi (1+\delta_{\ell\ell'}) C_\ell^{\rm data}
     C_{\ell'}^{\rm data}}{z_\ell^2 z_{\ell'}^2
     (H^L_{\ell\ell'})^{2} I^2} \nonumber \\
     &=& \frac{8\pi^3}{27} \frac{(1+\delta_{\ell\ell'}) C_\ell^{\rm
     data} C_{\ell'}^{\rm data}}{ \left(z_\ell z_{\ell'}
     H^L_{\ell\ell'} \right)^2 (\SNR)^2(N^{zz})^2} .
\end{eqnarray}
We then combine all the estimators
$(\reallywidehat{g_{LM}})_{\ell\ell'}$ with inverse-variance weighting
to obtain the minimum-variance estimator{\footnote{{The 
statistical independence of the different $\ell\ell'$ estimators is discussed further 
in the third-to-last paragraph.}}},
\begin{equation}
     \reallywidehat{g_{LM}} = \frac{\sum_{\ell\ell'} (\reallywidehat{g_{LM}})_{\ell
     \ell'}  (\Delta g_{LM})_{\ell\ell'}^{-2}}{
     \sum_{\ell\ell'} (\Delta g_{LM})_{\ell\ell'}^{-2}}.
\label{eqn:minvar}
\end{equation}
Note that the sums here are only over $\ell\ell'$ pairs that
have even $\ell+\ell'+L$, $|\ell-\ell'| \leq L \leq
\ell+\ell'$, and $\ell,\ell' \leq \ell_{\rm max}$.  Given the
reality of $z(\hat n)$, we sum only over
$\ell' \geq \ell$ to avoid double counting the contributions
from $A^{LM}_{\ell\ell'}$ and $A^{LM}_{\ell'\ell}$.
The variance $(\Delta g_{LM})^2$ with which $g_{LM}$ can be
measured is the inverse of the denominator; i.e.,
\begin{equation}
     (\Delta g_{LM})^{-2} = \frac{27}{16\pi^3} \sum_{\ell\ell'}
     \frac{ \left[H^L_{\ell\ell'} z_\ell z_{\ell'} (\SNR) N^{zz}
     \right]^2} {
     (C_\ell+N^{zz}) (C_{\ell'} +N^{zz})}.
\label{eqn:variance}     
\end{equation}
In this equation, the sum is now over {\it all}  $\ell$-$\ell'$ pairs
with $|\ell-\ell'|\leq L \leq \ell+\ell'$,
$\ell+\ell'+L$ even, and $\ell,\ell' \leq \ell_{\rm max}$,
which we obtain by using
$1+\delta_{\ell\ell'}=2$ for $\ell=\ell'$ and then including
both $\ell>\ell'$ and $\ell<\ell'$ and dividing by 2
\cite{Pullen:2007tu}.  We can then use
\begin{equation}
     C_\ell = \frac{3}{\pi}\sqrt{\frac{3}{2}} \frac{z_\ell^2}{2\ell+1} N^{zz}(\SNR),
\label{eqn:ClSNR}
\end{equation}
to obtain for the $\SNR\to \infty$ limit,
\begin{equation}
     (\Delta g_{LM})^{-2} =\! \frac{1}{8\pi} \sum_{\ell\ell'}
               (2\ell+1)(2\ell'+1) (H^L_{\ell\ell'})^2, \quad
               {\rm as}\,\,
               \SNR\to\infty.
\label{eqn:gLMmin}
\end{equation}
The smallest $g_{LM}$ that can be distinguished at the
$\sim3\sigma$ level from the null hypothesis $g_{LM}=0$ is then 
$ g_{LM,{\rm min}} \simeq 3 (\Delta g_{LM})$.

\section{Smallest detectable anisotropies}
\label{sec:smallest}

\subsection{Results for dipole anisotropy}

We now illustrate with the dipole $L=1$.  To do so, we note that
the only nonvanishing $\ell$-$\ell'$ pairs are those with
$\ell'=\ell\pm1$.  We then choose to take $\ell'=\ell+1$ and
multiply by two and use $(H^{L=1}_{\ell,\ell+1})^2 =
(\ell+3)(\ell-1)/[(\ell+1)(2\ell+1)(2\ell+3)]$ to obtain
\begin{eqnarray}
     g_{1M,{\rm min}} &=& 3 \left[ \sum_{\ell=2}^{\ell_{\rm max}-1}
     \frac{54}{\pi} \frac{ (\SNR)^2}{\ell^2 (\ell+1)^3
     (\ell+2)^2} \right. \nonumber \\
     & & \left. \times \frac{1}{ \left( 1 +{C_\ell}/{N^{zz}}
     \right)\left( 1 +{C_{\ell+1}}/{N^{zz}} \right)}
     \right]^{-1/2}.
\label{eqn:g1min}
\end{eqnarray}
We take the sum on $\ell$ up to $\ell_{\rm max}-1$ so that the
maximum $\ell'=\ell+1$ corresponds to the largest multipole
$\ell_{\rm max}$ that is measured.

\begin{figure}[htbp]
\centering
\hspace*{-0.1cm}\includegraphics[width=0.5\textwidth]{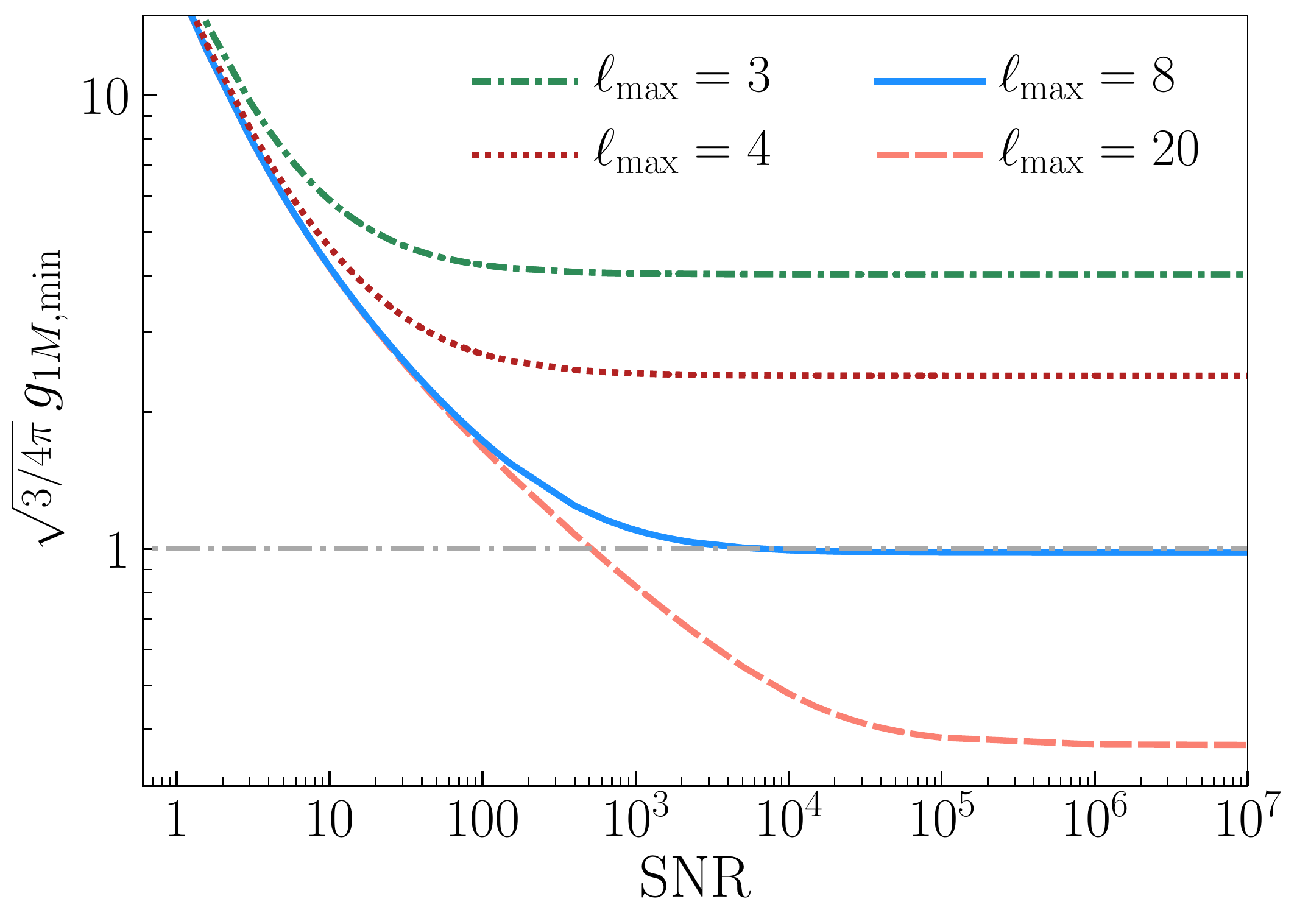}
\caption{The smallest detectable (at the $3\sigma$ level)
      dipole-anisotropy coefficient $g_{1M}$ (multiplied by
      $\sqrt{3/4\pi}$).  Results are shown
      as a function of the signal-to-noise ratio for the
      isotropic GW signal and for several values of the maximum
      timing-residual multipole moment $\ell_{\rm max}$ (which
      is $\ell_{\rm max}\simeq \sqrt{N_p}$ in terms of the number
      $N_p$ of pulsars).}
\label{fig:g1sensitivity}
\end{figure}

We then set $\ell_{\rm max} \simeq N_p^{1/2}$, where $N_p$ is
the number of pulsars, if these pulsars are distributed roughly
uniformly on the sky; the sensitivity to 
higher-$\ell$ modes will be exponentially reduced.  We then plot
in Fig.~\ref{fig:g1sensitivity} the
smallest detectable (at the $\gtrsim 3\sigma$ confidence level)
dipole-anisotropy amplitude $g_{1M}$ as a function of the
signal-to-noise ratio SNR with
which the isotropic signal is detected and for different
numbers of pulsars.  The results can be understood by noting
that Eq.~(\ref{eqn:g1min}) becomes, in the $\SNR\to\infty$
limit and in the limit $\ell_{\rm max} \gg 1$,
\begin{equation}
   g_{1M,{\rm min}} \sim \frac{6\sqrt{2\pi}}{\ell_{\rm max}}, \qquad {\rm
   as}\,~\SNR\to\infty.
\label{eqn:lownoiselimit}   
\end{equation}
However, this asymptotic limit is reached only for very large
SNR, given the very rapid decrease of $C_\ell/N^{zz}$ with
$\ell$.  In more physical terms, the anisotropy is inferred
through correlations between spherical-harmonic modes of
different $\ell$, and so individual modes of higher $\ell$ must
be measured with high signal-to-noise.  The steep dropoff of
$C_\ell$ with $\ell$ (each of the seven $\ell=3$ moments has a
signal-to-noise that is smaller by a factor of 5 than that for
each quadrupole moment) requires that the isotropic signal
(which is very heavily dominated by the quadrupole) be detected
with very high significance.  In the low-SNR limit,
Eq.~(\ref{eqn:g1min}) is approximated,
\begin{equation}
   g_{1M,{\rm min}} \sim \frac{28}{\SNR}, \qquad {\rm
   as}\,~\SNR\to0;
\label{eqn:highnoiselimit}   
\end{equation}
given the rapid decrease of the summand with $\ell$ in this
low-SNR limit, this result is obtained for any $\ell_{\rm max}
\geq3$.  In practice, this $\SNR\to0$ limit is somewhat
academic (and optimistic), as the factor $C_{\ell=2}/N^{zz}$ in
the denominator of the summand in Eq.~(\ref{eqn:g1min}) is
already 1.8 for $\SNR=3$.  Thus, the numerical result is a bit
larger, even for $\SNR=3$, than indicated by
Eq.~(\ref{eqn:highnoiselimit}).  The numerical results in
Fig.~\ref{fig:g1sensitivity} then indicate that the scaling with
higher SNR is more like $(\SNR)^{-1/2}$ rather than $(\SNR)^{-1}$ at
higher SNR, and that the $\SNR\to\infty$ limit (the
``pulsar-number--limited regime'') is achieved for ${\SNR} \gtrsim
1000$.  This can be understood by noting that, for example, the
$C_{\ell}/N^{zz}$ in the denominator of the summand in
Eq.~(\ref{eqn:g1min}) does not reach unity until the
signal-to-noise ratio grows, for $\ell=4$, to $\SNR\gtrsim 60$,
and for $\ell=8$, to $\SNR\gtrsim 1500$.  This then shows that
the benefit of $\gtrsim 16$ ($\gtrsim64$) pulsars for this
particular measurement is limited until the GW signal is
detected at ${\SNR}\gtrsim 60$ ($\gtrsim1500$).

If we wanted first to simply establish the existence of a
dipole, without specifying its direction, then our observable
would be the overall dipole amplitude,
\begin{equation}
     d = \sqrt{\frac{3}{4\pi}} \left[\sum_M |g_{1M}|^2
     \right]^{1/2},
\end{equation}
where we have included the factor of $\sqrt{4\pi}$ so that
$d\leq1$.  The SNR with which this can be established is then
$\sqrt{3}$ times that with which any individual $g_{1M}$ can be
measured, and so the smallest detectable (at $3\sigma$) dipole
has an amplitude
\begin{eqnarray}
     d_{\rm min} &\simeq& \frac{8}{\SNR}, \qquad
     \mathrm{as}\ \ \ \SNR\to0,
     \nonumber \\
     d_{\rm min} &\simeq& \frac{4}{\ell_{\rm max}}, \qquad
     \mathrm{as}\ \ \ \SNR\to\infty,
\end{eqnarray}
again noting that the $\SNR\to0$ limit is likely overly
optimistic for $\SNR \gtrsim 3$ and the $\SNR\to\infty$ limit is
valid for $\ell_{\rm max}\gg1$.

\subsection{Higher $L$ modes}

\begin{figure}[tp]
\centering
\includegraphics[width=0.5\textwidth]{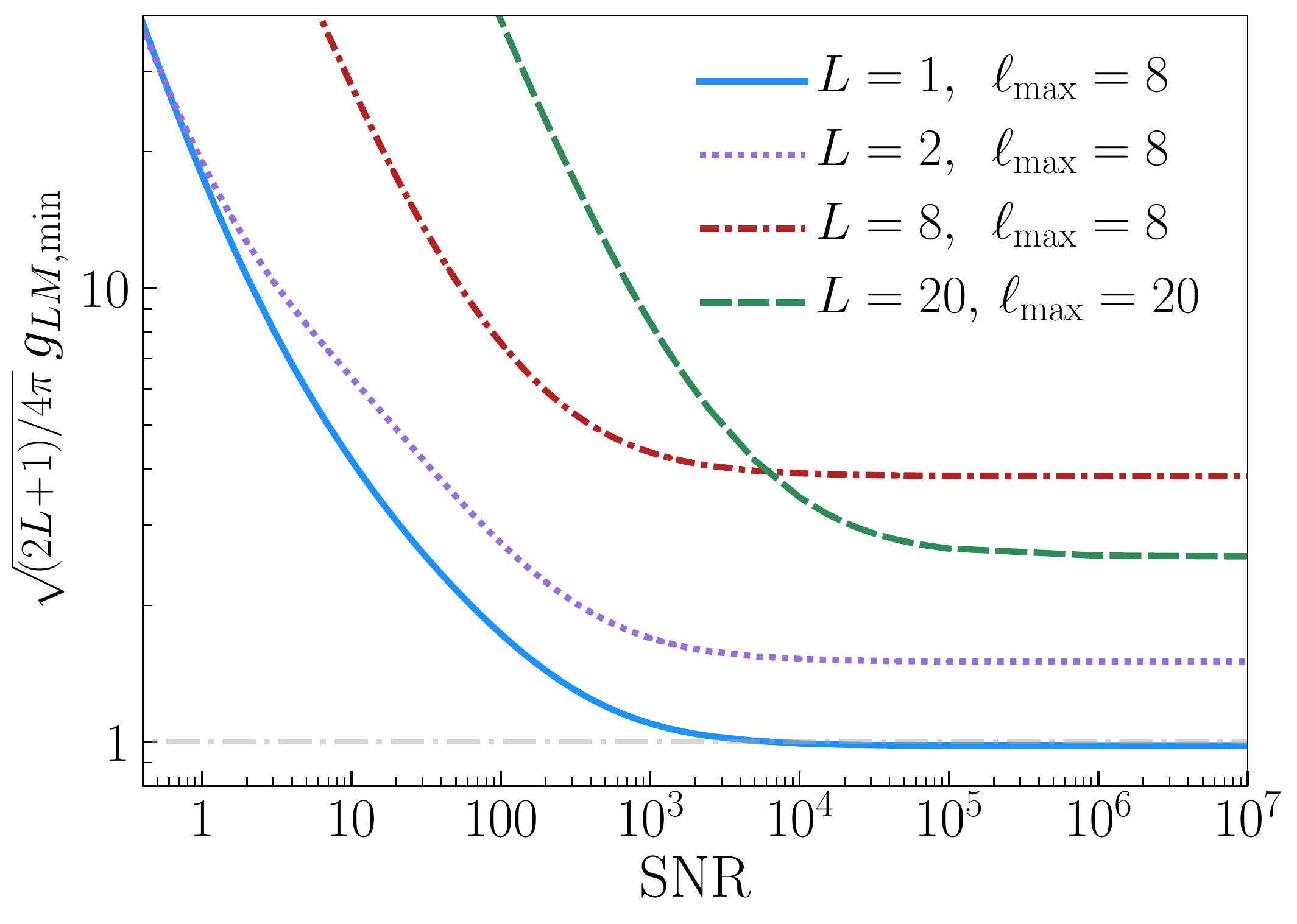}
\caption{The smallest detectable anisotropy coefficient $g_{LM}$,
     for $L=\{1,2,8,20\}$, as a function of the total SNR with which
     the isotropic GW signal is detected.  Results are provided
     for $\ell_{\rm max}=\{8,8,8,20\}$ respectively for the
     different $L$.}
\label{fig:higherL}
\end{figure}

The results for higher $L$ of the smallest detectable $g_{LM}$
are easily obtained by numerical evaluation of
Eq.~(\ref{eqn:gLMmin}) and shown for different $\ell_{\rm max}$
and (SNR) in Fig.~\ref{fig:higherL}.  The qualitative dependence
of the results are similar to those for $g_{1M,{\rm min}}$,
although the sensitivity to higher-$L$ anisotropies is reduced a
bit (e.g., by about 50\% for $L=5$) relative to the dipole sensitivity.

\subsection{A gravitational-wave beam}
\label{sec:beam}

\begin{figure}[bt]
\centering
\includegraphics[width=0.5\textwidth]{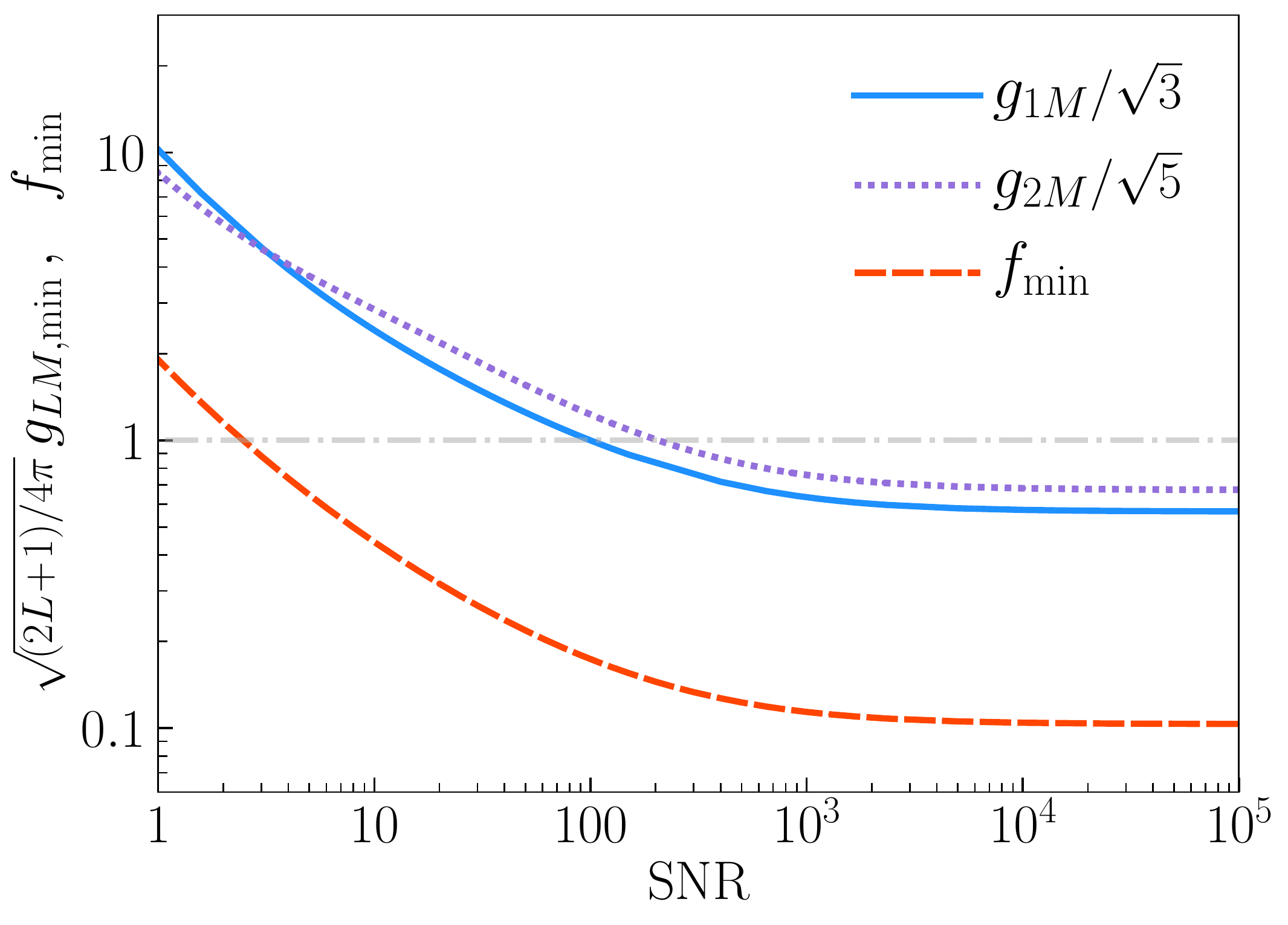}
\caption{For  $\ell_\mathrm{max}=8$, the smallest detectable dipole-anisotropy amplitude
     $d$ (which is $3^{-1/2}$ times the $g_{1M,\mathrm{min}}$
     plotted in Fig.~\protect\ref{fig:g1sensitivity}), and the smallest detectable 
     quadrupole-anisotropy amplitude, shown
     together with the smallest detectable (again, at $3\sigma$)
     beam amplitude $f$ obtained with measurements of $g_{L0}$
     up to L=8 is $f_{\mathrm{min}}\simeq0.1$ in the high SNR
     limit.}
\label{fig:minVsummed}
\end{figure}

Suppose that a gravitational-wave signal has been detected and
that we wish to determine the fraction of the local
gravitational-wave energy density coming from a specific
direction.  To be more precise suppose that we model the
gravitational-wave signal as an isotropic uncorrelated
background plus a flux of gravitational waves all coming from
some specific direction (e.g., the direction of some specific
SMBH binary candidate), which we take to be in the $\hat z$
direction, that makes up a fraction $f$ of the local GW energy
density.  This situation is described by anisotropy coefficients
$g_{LM} = \sqrt{4\pi}  f  \sqrt{2L+1}\delta_{M0}$.  The minimum-variance
estimator for the amplitude $f$ is then obtained by summing the
minimum-variance estimators for $g_{L0}$ (scaled by
$\sqrt{4\pi} $), with inverse-variance weighting.  In Fig.~\ref{fig:minVsummed}, 
we plot the smallest $f$ using the results above for $g_{LM,{\rm min}}$ for 
$L\leq 8$, detectable with measurements of $g_{LM}$
up to $L=8$, as a function of SNR from a single map and 
find it approaches $f_{\rm min} \simeq 0.1$ in the $\SNR\gtrsim1000$
regime. {It may be possible, however, to improve the sensitivity 
to a specific gravitational-wave point source if the signal is characterized 
by more than the incoherent flux assumed here.}

\section{Multiple maps}
\label{sec:multiplemaps}

So far, we have assumed that there is a single timing-residual map
$z(\hat n)$ obtained by convolving the time-domain data with a
single window function.  Suppose, however, that the time-domain
data are convolved with $n_w$ different time-domain window
functions that have negligible overlap in frequency space (or in
phase).  For example, if we were to have measurements performed,
every two weeks for $\sim 10$ years, yielding $\sim250$
measurements for each pulsar,  the time-domain window functions
could be taken to be the $\sim250$ different time-domain Fourier
modes.  In this case, we will have $n_w\sim250$ statistically
independent timing-residual maps $z^i(\hat n)$, with
$i=1,2,\ldots,n_w$.  If the Hellings-Downs power spectrum is
detected with signal-to-noise ratio $(\SNR)_i$ in each individual
map $i$, then the signal-to-noise ratio (squared) for the entire
experiment, after co-adding all the information, will be $(\SNR)^2
= \sum_i (\SNR)_i^2$.

The optimal estimator for any given $g_{LM}$ is then obtained by
adding (with inverse-variance weighting) the estimators
$\reallywidehat{g_{LM}^i}$ from each map $i$; i.e., we augment
Eq.~(\ref{eqn:variance}) with an additional sum over $i$ and
replace the SNR, the power spectrum $C_\ell$, and noise power
spectrum $N^{zz}$ by those---$(\SNR)_i$, $C_\ell^i$, and
$N_i^{zz}$---associated with the $i$th map:
\begin{equation}
     (\Delta g_{LM})^{-2} = \frac{27}{16\pi^3} \sum_i \sum_{\ell\ell'}
     \frac{ (H^L_{\ell\ell'} z_\ell z_{\ell'} (\SNR)_i N_i^{zz})^2} {
     (C_\ell^i+N_i^{zz}) (C_{\ell'}^i +N_i^{zz})}.
\label{eqn:variancei}     
\end{equation}
The SNR, power spectrum, and noise power spectrum for each map
are related by
\begin{equation}
     C^i_\ell = \frac{3}{\pi}\sqrt{\frac{3}{2}} \frac{z_\ell^2}{2\ell+1} N^{zz}_i(\SNR)_i,
\label{eqn:ClSNR2}
\end{equation}

\begin{figure}[bt]
\centering
\includegraphics[width=0.5\textwidth]{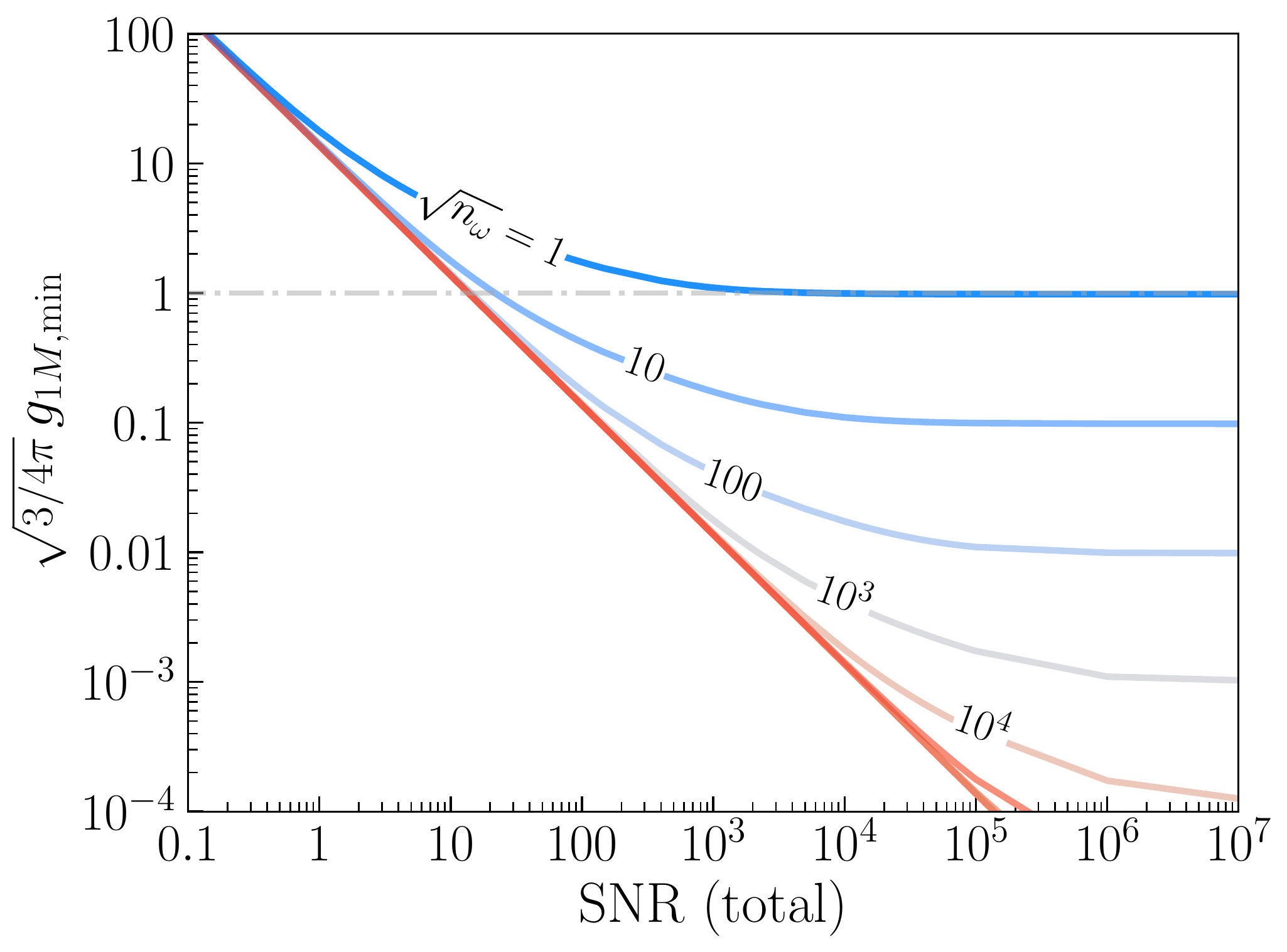}
\caption{The smallest detectable dipole coefficient $g_{1M,{\rm
    min}}$ as a function of the {\it total} signal-to-noise for
    $\ell_{\rm max}=8$.  The different curves show results
    obtained for different numbers of statistically independent
    maps {$n_\omega$}, {\it assuming that the total SNR is distributed equally
    among all these maps}.}
\label{fig:multiplemaps}
\end{figure}

In the $\SNR\to0$ limit, these replacements result (given
$\sum_i (\SNR)_i^2 = (\SNR)^2$) in the same anisotropy sensitivity as inferred for a single map in
this limit in Eq.~(\ref{eqn:highnoiselimit}).  If, however, the signal-to-noise ratio $\SNR_i$ in
some number $n_{\rm high}$ of maps is high enough (e.g., $(\SNR)_i
\gtrsim 60$ for $N_p=16$ or $(\SNR)_i \gtrsim 1500$ for $N_p=64$) 
that  pulsar-number--limited regime is reached {\it in each
individual map}, then the sensitivity to anisotropy can be
improved by a factor $\sqrt{n_{\rm high}}$ relative to that,
Eq.~(\ref{eqn:lownoiselimit}), as shown in
Fig.~\ref{fig:multiplemaps}.  It must be kept in mind that the
improvement shown in Fig.~\ref{fig:multiplemaps} possible with
additional maps is achieved {\it only if the total SNR is split
evenly among all of these many maps}.

The remaining question, then, is how the total signal-to-noise is
distributed among the maps.  In the best-case scenario, it will
be distributed equally among the $n_w$ maps.  If so, then
sensitivity to anisotropy could be improved, in principle, by the
factor $\sqrt{n_w}$ over that in Eq.~(\ref{eqn:lownoiselimit}),
as shown in Fig.~\ref{fig:multiplemaps}.
This improvement would require, however, that the {\it total}
signal-to-noise be $\sim \sqrt{n_w}$ larger than that
($\SNR\gtrsim 60$ for $N_p=16$ and $\SNR\gtrsim 1500$ for
$N_p\gtrsim 64$) for a single map.

Given the likely (given the most promising astrophysical
scenarios) decrease of the signal with GW frequency, however,
the signal-to-noise will probably be dominated by a small subset
of the maps (those at the lowest frequencies).  If so, then
$n_{\rm high}$ may be far smaller than $n_w$, and the
sensitivity, from multiple maps, to anisotropy
will be only marginally improved over the single-map estimate in
Eq.~(\ref{eqn:lownoiselimit}).

\section{Discussion}
\label{sec:discussion}

We have discussed the search for anisotropy in a PTA-detected
gravitational-wave signal in terms of bipolar spherical
harmonics for idealized measurements parametrized in terms of
the number of pulsars (assumed to be uniformly distributed on
the sky) and the signal-to-noise ratio (SNR) with which the
isotropic signal is established.  We focussed our attention
first on the case of a single timing-residual map $z(\hat n)$
(obtained from the convolution of the data with a single
time-domain window function) and then discussed the
generalization to multiple maps (which take into account 
more of the time-domain information).

We considered a search for anisotropy in an
uncorrelated and unpolarized GW background in which the
anisotropy is independent of GW frequency.  In this case, the
anisotropy is parametrized entirely in terms of spherical-harmonic
coefficients $g_{LM}$.  We derived the optimal estimators for
these $g_{LM}$ for idealized measurements in which $N_p$ pulsars
are distributed roughly uniformly on the sky and the
same timing-residual noise in each pulsar.  We then obtain
the variance with which each $g_{LM}$ can be determined; this
variance is expressed in terms of the signal-to-noise with which
the isotropic signal is detected and in terms of the number of
pulsars.

The main qualitative upshot of the analysis is that the
isotropic signal will have to be established very well before
there is any possibility to detect anisotropy.  The reason stems
from the the fact that the anisotropy is obtained (for odd $L$) through
cross-correlation of spherical-harmonic modes $z_{\ell m}$ of
the timing-residual map of different $\ell$ and from the fact
(inferred from Eq.~(\ref{eqn:SNR})) that $94\%$ of the $(\SNR)^2$
for the isotropic signal comes from $\ell=2$, with only $6\%$
coming from higher modes.  Our numerical results in
Fig.~\ref{fig:g1sensitivity} show that with
a single map it will require the isotropic signal to be
established with $\SNR\gtrsim1000$ before even the maximal
dipole anisotropy can be distinguished, at the $3\sigma$ level,
from a statistically isotropic background.  This would,
moreover, require $\gtrsim60$ pulsars spread uniformly over the
sky.  The sensitivity to a dipole signal can be improved with
more pulsars and/or (as Fig.~\ref{fig:multiplemaps} shows) with
multiple maps, constructed with different
statistically-independent time-domain window functions.  This
latter improvement can be achieved, however, only if
the signal-to-noise is spread evenly among these different
maps.  Fig.~\ref{fig:higherL} indicate the additional challenge
facing a search for higher-order anisotropy.

When discussing the prospects to detect anisotropy, we must be
careful to state clearly the question we are trying to answer.
Here we have focused on the sensitivity to departures from {\it
statistical isotropy} parametrized in terms of
spherical-harmonic coefficients $g_{LM}$, under the null
hypothesis of a statistically-isotropic background.  This
sensitivity is limited not only by measurement noise, but also
by cosmic variance.  In our null hypothesis of a statistically
isotropic signal, the spherical-harmonic coefficients $z_{\ell
m}$ for the map are selected, in the limit of no-noise
measurements, from a distribution with variance $C_\ell$.
Departures from statistical isotropy show up, roughly speaking,
in terms of disparities between the amplitudes of the different
$m$ modes for a given $\ell$.  The conclusion of our analysis is
that this is difficult to establish given the variance $C_\ell$
under the null hypothesis.

A measurement that is consistent with {\it statistical isotropy}
may still well exhibit some evidence that the local GW
background is a realization that exhibits anisotropy.
Suppose, for example, that we had precise measurement of the
five timing-residual quadrupole moments $z_{2m}$ and found that
the $z_{22}$ and $z_{2,-2}$ components were significantly larger
than the other three quadrupole moments.  Our calculation
[obtained by evaluating
Eq.~(\ref{eqn:gLMmin}) with only the $\ell=\ell'=2$ term]
indicates that it would be impossible to infer from this
measurement any departure from statistical isotropy.  Still, if
such a result were observed, it would indicate that the local GW
signal is coming primarily from the $\pm \hat z$ direction.
If there was indeed a strong candidate GW source (e.g., a
SMBH-binary candidate) in the $\hat z$ direction, then
this observation would provide some evidence that the GW signal
was coming predominantly from that source.

Our initial calculations explored the detectability of
anisotropy from a single timing-residual map obtained by
convolving the data with a single time-domain window function.
If, however, multiple maps that explore different GW frequency
ranges can be obtained, then there are prospects to co-add the
anisotropy estimators from those maps to improve upon the
pulsar-number limit that arises from a single map.  Significant
improvement in this way requires, however, that the $z_{\ell m}$
are measured with high SNR in multiple maps.

We also considered the prospects to measure the fraction $f$ of
the local GW intensity that comes from a given direction.  We
conclude also that measurement of $f$ will be similarly
challenging:  For example, we found a value $f_{\rm
min}\simeq0.1$ for the smallest detectable fraction for a survey
with $64$ pulsars with $\SNR\gtrsim 1000$ for the isotropic
signal.  This calculation leaves out ingredients (e.g.,
timing, polarization, and source evolution) that, if included
in the analysis, might improve the ability to localize a point
source.  Still, the rough conclusions and scalings of the
sensitivities with SNR and pulsar number should translate to
those for a more complete point-source search.

We also note, for possible comparison with previous work in
configuration space
\cite{Anholm:2008wy,Mingarelli:2013dsa,Gair:2014rwa},
that a sky described by BiPoSHs $\ALM$ has a two-point
correlation function,
\begin{equation}
     C(\hat n, \hat m) = C(\Theta) + \sum_{\ell \ell' L M} \,\ALM
     \,\,\{Y_\ell(\hat n)\otimes Y_{\ell'}(\hat m)\}_{LM},
\label{eqn:generalcorrelation}     
\end{equation}
where 
\begin{equation}
     \{Y_\ell(\hat n)\otimes Y_{\ell'}(\hat m)\}_{LM} =
     \sum_{mm'} \cleb{L}{M}{\ell}{m}{\ell'}{m'} \, Y_{\ell
     m}(\hat n) Y_{\ell'm'}(\hat m),
\label{eqn:biposhs}
\end{equation}
are the bipolar spherical harmonics (BipoSHs).  These BiPoSHs
constitute a complete orthonormal basis for functions of $\hat n$
and $\hat m$ in terms of total-angular-momentum states labeled
by quantum numbers $L$ and $M$ composed of angular momentum
states with $l$ and $l'$; they are an alternative to the outer
product of the $\{l,m\}$ and $\{l',m'\}$ bases. {It should be 
possible to identify these bipolar spherical harmonics to the anisotropic 
correlation functions worked out in Refs.~\citep{Mingarelli:2013dsa,Gair:2014rwa,
Taylor:2015udp}, but we leave this exercise to future work.}

The analysis presented here should be straightforwardly
generalized to astrometric GW searches.  As shown in
Ref.~\cite{Qin:2018yhy}, the E-mode map from an astrometry
survey provides the
same information as a timing-residual map.  Therefore, everything
said here about a timing-residual map can be applied equally to the
E-mode map.  The higher
density of stellar astrometric sources on the sky may ultimately
allow higher $\ell_{\rm max}$ but the advantage of this higher
$\ell_{\rm max}$ for anisotropy searches can be capitalized upon
only with a sufficiently high SNR.  
The B modes in the astrometry map can provide additional
statistically independent information and, when combined with
the E modes and/or timing residuals, can conceivably improve the
sensitivity to anisotropy by a factor of $\sqrt{2}$.

The numerical results we find for the sensitivity to anisotropy
may be optimistic, given the idealizations assumed
here.  Uneven distribution of pulsars on the sky and/or
pulsar-to-pulsar variations in the timing-residual noises will
degrade the sensitivity.  There is another, more subtle,
caveat:  The estimator in Eq.~(\ref{eqn:minvar}), and the
expression, Eq.~(\ref{eqn:variance}), for its variance, are
derived under the assumption that the different
$\reallywidehat{A^{LM}_{\ell\ell'}}$ are statistically independent.
Although the covariance between any two different
$\reallywidehat{A^{LM}_{\ell\ell'}}$ vanishes (except for those with
$\ell\leftrightarrow \ell'$), they are {\it not} statistically
independent.  The variance will thus most generally be a bit
larger, and the sensitivity to anisotropy a bit degraded.  We
anticipate, though, that for the low-$L$ values considered here
that this will be a relatively small (perhaps $\sim10\%$)
effect, although this should be evaluated further with Monte
Carlo simulations of isotropic GW signals.

We hope that the approach developed here provides a conceptually
straightforward way to understand the search for anisotropy in
the GW background and aids in the development of
observational/analysis strategies for the PTA search for
gravitational waves. It will be interesting in future work to
compare the results here to those obtained from detailed
simulations of the PTA analysis pipeline, as well as with those inferred 
from a fully Bayesian approach 
(see similar applications for the cosmic microwave background~\cite{Das:2015gca,Shaikh:2019dvb}, for example). 
It will also be interesing to extend the analysis here to seek anisotropies in
the polarization of the GW background, as parametrized, for
example, by GW Stokes parameters \cite{Conneely:2018wis}, or
anisotropies in the frequency dependence of the GW background. {In the former work, authors compare 
the statistics of the GW Stokes parameters (spin 4) and concentrate on the isotropic spectra and cross-spectra, 
similar to this work, which builds a bridge between the two formalisms in terms of the statistical anisotropy 
of the amplitude fields that, as mentioned above, is a generalization of the power anisotropy.}

{In this paper we have considered ideal measurements in which 
GW-induced redshifts are measured as a function of position on the sky (with 
uniform sensitivity over the entire sky) and as a function of time (with uniform 
sampling/sensitivity).  In this case, the harmonic-space basis is a cross product 
of spherical harmonics (for the sky) and Fourier modes for the time domain.  
In this idealized case, each (spherical-harmonic)--(time-domain Fourier mode) 
is statistically independent, for the GW background we are considering (i.e., that 
specified in Eq.~\eqref{eqn:generalized}).  In practice, incomplete/irregular sky 
coverage, nonuniform timing-residual noise, and irregularities in the observation 
times destroy this elegant diagonalization.  Techniques have been developed to 
deal with the cross-correlations induced by these imperfections on the idealized 
eigenmodes.  For example, one can deal with real-space correlations, as done in 
prior work (e.g., Refs.~\citep{Mingarelli:2013dsa,Gair:2014rwa,Taylor:2015udp}).  
Another option is to work with experiment-specific signal-to-noise eigenmodes 
(e.g., as being developed in Ref.~\citep{TCY:2019}).  Most generally, these 
imperfections will reduce the sensitivity to isotropic signals and/or anisotropy in
 the signal relative to those obtained here, assuming ideal measurements, 
 although more detailed specification of the experiment is required to evaluate 
 precisely the reduction in sensitivity.}
 
{During the preparation of this work, we learned of related work \cite{TCY:2019} 
in preparation that addresses prospects to detect anisotropy with a focus on 
developing a formalism to produce maps of the gravitational wave background
from pulsar timing array measurements.  We plan to follow up with detailed 
comparison of that work and the formalism described here.}

\begin{acknowledgments}
We thank K.\ Boddy, L.\ Kelley, C.\ Mingarelli, and T.\ Smith
for useful discussions.  SCH acknowledges the
support of a visitor grant from the
New-College-Oxford/Johns-Hopkins Centre for Cosmological
Studies and Imperial College President's Scholarship. AHJ acknowledge support from STFC in the UK. This work was supported at Johns Hopkins by NASA Grant
No.\ NNX17AK38G, NSF Grant No.\ 1818899, and the Simons Foundation.
\end{acknowledgments}

\vspace*{-0.5cm}

\section{Erratum}

In the original version of this paper, the equality $g_{LM}=\sqrt{4\pi}f\sqrt{2L+1}\delta_{M0}$ was missing the factor $2L+1$ on the right hand side. We thank Yacine Ali-Ha\"{i}moud, Tristan Smith and Chiara Mingarelli for spotting this error. As a result, the orange dashed curve in Fig.~\ref{fig:minVsummed} was reduced by a factor of $\sim2$ and our estimates for the $f_{\rm min}$ changed from $0.28$ to $0.1$ in the high SNR limit.

\end{document}